\documentclass[manuscript]{biometrika}

\usepackage{amsmath,amssymb}

\usepackage{times}
\usepackage{bm}
\usepackage{natbib}
\usepackage{mathbbol}
\usepackage{caption}

\usepackage[plain,noend]{algorithm2e}

\makeatletter
\renewcommand{\algocf@captiontext}[2]{#1\algocf@typo. \AlCapFnt{}#2} 
\def\@algocf@capt@plain{top}
\renewcommand{\algocf@makecaption}[2]{%
	\addtolength{\hsize}{\algomargin}%
	\sbox\@tempboxa{\algocf@captiontext{#1}{#2}}%
	\ifdim\wd\@tempboxa >\hsize
	\hskip .5\algomargin%
	\parbox[t]{\hsize}{\algocf@captiontext{#1}{#2}}
	\else%
	\global\@minipagefalse%
	\hbox to\hsize{\box\@tempboxa}
	\fi%
	\addtolength{\hsize}{-\algomargin}%
}
\makeatother

\usepackage{color}
\usepackage{array}
\usepackage{verbatim}
\usepackage{mathrsfs}
\usepackage{multirow}
\usepackage{amsfonts}
\usepackage{graphicx}
\usepackage{esint}
\usepackage{epstopdf}
\usepackage{xargs}[2008/03/08]
\usepackage{xr}
\usepackage{soul}

\usepackage{tikz}
\usetikzlibrary{shadings,intersections}
\definecolor{mydarkgreen}{rgb}{0.101961, 0.478431, 0.101961}
\definecolor{mydarkred}{rgb}{0.647059, 0.164706, 0.164706}
\definecolor{fillcolor}{rgb}{0.623529, 0.968627, 0.623529}

\ifx\du\undefined
\newlength{\du}
\fi
\providecommand{\tabularnewline}{\\}

\newenvironment{lyxlist}[1]
{\begin{list}{}
		{\settowidth{\labelwidth}{#1}
			\setlength{\leftmargin}{\labelwidth}
			\addtolength{\leftmargin}{\labelsep}
			}}
	{\end{list}}

\usepackage{ifpdf} 
\ifpdf 

\IfFileExists{lmodern.sty}{\usepackage{lmodern}}{}

\fi 

\usepackage{enumitem}

\usepackage{etoolbox}
\makeatletter
\patchcmd{\@makecaption}
{\parbox}
{\advance\@tempdima-\fontdimen2} 
{}{}
\makeatother

\makeatletter
\newsavebox{\@brx}
\newcommand{\llangle}[1][]{\savebox{\@brx}{\(\m@th{#1\langle}\)}%
	\mathopen{\copy\@brx\kern-0.5\wd\@brx\usebox{\@brx}}}
\newcommand{\rrangle}[1][]{\savebox{\@brx}{\(\m@th{#1\rangle}\)}%
	\mathclose{\copy\@brx\kern-0.5\wd\@brx\usebox{\@brx}}}
\makeatother

\global\long\def\expect{{E}}

\global\long\def\bW{{W}}
\global\long\def\bX{Q}
\global\long\def\bx{\xi}

\global\long\def\real{\mathbb{R}}
\global\long\def\hpca{h_{pca}}
\global\long\def\hreg{h_{reg}}
\global\long\def\hpcad#1{h_{pca}^{#1}}
\global\long\def\Op{O_{P}}
\global\long\def\manifold{\mathcal{M}}
\global\long\def\txm{T_{x}\mathcal{M}}

\global\long\def\hilbert{\mathcal{\mathcal{L}}^{2}}
\global\long\def\covarop{\mathcal{C}}
\newcommandx\estneighbor[2][usedefault, addprefix=\global, 1=\delta, 2=x]{\hat{\mathcal{N}}_{\mathcal{L}^{2}}(#1,#2)}
\newcommandx\estneighborM[2][usedefault, addprefix=\global, 1=\delta, 2=x]{\mathcal{N}_{\manifold}(#1,#2)}
\newcommandx\estneighborindex[2][usedefault, addprefix=\global, 1=\delta, 2=x]{\hat{\mathcal{I}}(#1,#2)}
\newcommandx\ambientball[1][usedefault, addprefix=\global, 1=\delta]{\mathbb{B}_{#1}^{\mathcal{L}^{2}}(x)}

\global\long\def\asympgt{\gtrsim}

\global\long\def\ltwospace{\mathcal{L}^{2}(D)}

\newcommandx\manball[1][usedefault, addprefix=\global, 1=\delta]{\mathbb{B}_{#1}^{\manifold}(x)}
\global\long\def\diffop{\mathrm{d}}

\global\long\def\ltwo{\mathcal{L}^{2}}
\global\long\def\observed{\mathscr{X}}

\usepackage{setspace}
\usepackage{pdfpages}

\begin{document}
	
	\setstretch{1.5}
	
	\jname{Biometrika}
	\jyear{xxxx}
	\jvol{xxx}
	\jnum{x}
	\accessdate{xxxx}
	
	
	\markboth{Lin and Yao}{Functional Regression on Manifold}
	
	\title{Functional Regression on Manifold with Contamination}
	
	\author{Zhenhua Lin}
	\affil{Department of Statistics and Applied Probability, National University of Singapore, 117546, Singapore \email{stalz@nus.edu.sg}}
	
	\author{\and Fang Yao}
	\affil{Department of Probability and Statistics, School of Mathematical Sciences, Center for Statistical Science, Peking University, Beijing, 100871, China
		\email{fyao@math.pku.edu.cn}}
	
	\maketitle
	
	\begin{abstract}
		We propose a new method for  
		functional nonparametric regression with a predictor that resides on a finite-dimensional manifold but is only observable in an infinite-dimensional  space. Contamination of the predictor due to  discrete/noisy measurements is also accounted for.  By using functional local linear manifold
		smoothing, the proposed estimator enjoys a polynomial rate of convergence that adapts to the intrinsic manifold dimension and the contamination level. This is 
		in contrast to the logarithmic convergence rate in the literature of functional nonparametric regression.  We also observe a phase transition phenomenon regarding the interplay of the manifold dimension and the contamination level.
		We demonstrate that the proposed method has favorable numerical performance relative to commonly used methods via simulated and real data examples.
	\end{abstract}
	
	
	\section{Introduction}
	
	Regression with a functional predictor is of central importance
	in the field of functional data analysis, the field that has been
	advanced by \cite{Ramsay1997,Ramsay2002} and many others. Early development of functional regression
	focuses on functional linear models \citep{Cardot1999,Yao2005,Yuan2010}.
	Extensions of linear models include generalized linear regression \citep{Cardot2005,Mueller2005}, additive models \citep{Mueller2008}, quadratic models \citep{Yao2010}, among others. 
	These works prescribe specific forms of the regression model, and are regarded as functional parametric regression models \citep{Ferraty2006} 
	that entail efficient estimation procedures
	 and hence are well studied in the literature. 
	
	In contrast, functional nonparametric regression that does not impose structural constraints on  the regression function has received less attention. The first landmark development of nonparametric functional data analysis is the monograph of \citet{Ferraty2006}.  Recent
	advances in this direction include the Nadaraya--Watson
	estimator \citep{Ferraty2012} and the $k$-nearest-neighbor estimator \citep{Kudraszow2013}. The development of functional nonparametric
	regression is hindered by a theoretical barrier that is formulated
	in \citet{Mas2012} and is linked to the small ball probability
	problem \citep{Delaigle2010}. Essentially, in a rather general setting, the minimax rate of nonparametric
	regression on a generic functional space is slower than any polynomial
	of the sample size, which differs markedly from the polynomial minimax rates for many functional parametric regression procedures \citep[e.g.][for functional linear regression]{Hall2007,Yuan2010}. These endeavors on functional nonparametric
	regression do not exploit the intrinsic structure that is common
	in practice. For instance, \citet{Chen2012} suggested that functional
	data often possess a low-dimensional manifold structure which can be utilized for more efficient representation. By contrast, we exploit 
	the nonlinear low-dimensional	structure for functional nonparametric regression.
	
	Our method, which we call functional regression on manifold, assumes the model 
	\begin{equation}
	Y=g(X)+\varepsilon,\label{eq:nonparametric-functional-model}
	\end{equation}
	where $Y$ is a scalar response, $X$ is a functional predictor
	sampled from an unknown manifold $\manifold$, $\varepsilon$ is the
	error term independent of $X$, and $g$ is some unknown functional
	 to be estimated. In reality, the functional predictor $X$ is rarely fully observed. To accommodate this common practice, we assume that $X$ is recorded at a grid of points with noise. 
	The model \eqref{eq:nonparametric-functional-model} features a manifold structure $\manifold$
	that underlies the functional predictor $X$ and is assumed to be a finite-dimensional but potentially nonlinear submanifold of the function space $\ltwospace$, the space of square integrable
	functions defined on a compact domain $D\subset\real$. For a background on
	both finite-dimensional and infinite-dimensional manifolds, we refer readers to \citet{Lang1995} and \citet{Lang1999}.
	
	Data analysis with a manifold
	structure has been extensively studied in the statistical literature. For example, techniques have been invented to learn an unknown manifold based on a point cloud, such as locally linear embedding \citep{Roweis2000,Wu2018}, isomap \citep{Tenenbaum2000}, 
	t-SNE \citep{Maaten2008}, among many others.  
	Supervised learning on an unknown manifold has also been investigated, such as estimation of functions defined on a manifold \citep{Aswani2011,Cheng2013,Sober2019} and estimation of the gradient of such functions \citep{Mukherjee2010}. In addition, data analysis on a known manifold has been studied, such as fundamentals related to the Fr\'echet mean \citep{Bhattacharya2003,Bhattacharya2005,Bhattacharya2017}, manifold-valued function estimation  \citep{Yuan2012,Lin2016,Cornea2017,Lin2019}, manifold-valued principal component analysis \citep{Huckemann2010,Panaretos2014}, classification on manifolds \citep{Yao2019+}, and nonparametric manifold-valued inference \citep{Patrangenaru2015}.
	
	However, the literature specifically relating functional data to manifolds is scarce. \citet{Zhou2014}
	investigated functional principal component analysis on an irregular domain.  \citet{Chen2012} and  \citet{Lila2017} considered the representation and principal component analysis of functional 	data sampled from a manifold. Manifold-valued random functions were studied by \cite{Su2014}, \cite{dai:17:1} and \cite{Lin2019+}. 
	To the best of our knowledge,
	we are the first to consider a manifold structure in functional regression where a {global}
	representation of the low-dimensional functional predictor $X$ can be inefficient. 
	{For illustration, Example 1 in Supplementary Material exhibits a  random process taking values in a {one-dimensional}
		 submanifold of $\ltwo([0,1])$ while having an {{infinite}}
		number of components in its Karhunen--Lo\`{e}ve expansion}.

	When estimating the regression functional $g$ in \eqref{eq:nonparametric-functional-model}, we explicitly account for the hidden manifold structure by estimating the tangent spaces of the manifold. 
		Specifically, we first recover the observed functional predictors from their discrete/noisy measurements, and then adopt the local linear manifold smoothing \citep{Cheng2013}. While our approach and the one of \citet{Cheng2013} share the same intrinsic manifold setup, they fundamentally differ in  the ambient aspect, which raises challenging issues unique to functional data.  First, functional data 
		naturally live in an {infinite-dimensional} ambient space, while the Euclidean data considered by \citet{Cheng2013} have a finite ambient dimension.
	{Second, the effect of noise/sampling in the observed functional data needs to be explicitly treated, since functional data are discretely and noisily recorded in practice, which then introduces contamination of the functional predictor. This contamination issue is not encountered in \citet{Cheng2013}, or is only considered for linear regression of multivariate data \citep{Aswani2011,Loh2012}. Moreover, the contamination has an intrinsic dimension that grows with the sample size and thus is coupled with the ambiently infinite dimensionality. 
	
	The main contributions of this article are as follows. First, by
	exploiting structural information of the predictor, our proposal  
	entails an effective estimation procedure that adapts to the unknown manifold structure and the contamination level while
	maintains the flexibility of functional nonparametric regression. 
	 Second, by careful theoretical analysis, we confirm that the regression functional $g$ can be estimated at a polynomial convergence rate of the sample size, especially when only the contaminated functional predictors are available. This provides a new angle to functional nonparametric regression that 
	 is subject to a logarithmic rate \citep{Mas2012}.  Third, the contamination on predictors is  explicitly treated and is shown to be an integrated part of the convergence rate, which has not
	been well studied even in classical functional linear regression \citep{Hall2007}. 
	Finally, we discover that, the polynomial convergence
	rate exhibits a phase transition phenomenon, depending on the interplay
	between the manifold dimension and the contamination level. This type of phase transition has not yet been discovered in functional regression, and shares at least the same importance of those concerning estimation of mean/covariance functions \citep[e.g.][]{Cai2011, Zhang2016}. In addition, during our theoretical development, we obtain some results that are generally useful with their own merit, such as the consistency of the estimated intrinsic dimension and tangent spaces of the manifold in the presence of contamination.

	\section{Estimation of Functional Regression on Manifold \label{sec:general-theory}}
	
	\subsection{Step I: Recovery of Functional Data}\label{subsec:stepI}
	We assume that each predictor $X_{i}$ is observed at $m_{i}$ design points $T_{i1},\ldots,T_{im_{i}}\in D$. 
	Denote the observed value at $T_{ij}$ by $X_{ij}^{\ast}=X_{i}(T_{ij})+\zeta_{ij}$, where
	$\zeta_{ij}$ is random noise with mean zero and is independent of all $X_{i}$ and $T_{ij}$. The collection  ${\mathbb{X}}_{i}=\{(T_{i1},X_{i1}^{\ast}),\ldots,(T_{im_i},X_{im_{i}}^{\ast})\}$ represents all measurements for the realization $X_i$, and $\{\mathbb{X}_{1},\ldots,\mathbb{X}_{n}\}$
	constitutes the observed data for the predictor.  We shall clarify that, although each trajectory $X_i$ as a whole function resides on the manifold $\manifold$, the $m_i$-dimensional vector $\mathbb{V}_i=(X_i(T_{i1}),\ldots,X_i(T_{im_i}))$ does not. Consequently, the manifold assumption in \citet{Cheng2013} is violated for $\mathbb{V}_i$. 
	
	When $\inf_i m_i$ is sufficiently large or grows with the sample size, a scenario commonly referred to as the dense design, we may recover each function $X_{i}$
	based on the observed data $\mathbb{X}_{i}$ by individual smoothing estimation. 
	Popular smoothing techniques include the local linear smoother \citep{Fan1993} and 
	spline smoothing \citep{Ramsay2005}, among others. By applying one of these methods, we 
	obtain an estimate $\hat{X}_{i}$ of $X_{i}$, referred to as the contaminated version of $X_i$ that is used in the subsequent steps to estimate $g$. To be specific, we consider the local linear estimate of $X_{i}(t)$ given by $\hat{b}_{1}$ with 
	\begin{equation}
	(\hat{b}_{1},\hat{b}_{2})=\underset{(b_{1},b_{2})\in\real^{2}}{\arg\min}\frac{1}{m_i}\sum_{j=1}^{m_{i}}\left\{X_{ij}^{\ast}-b_{1}-b_{2}(T_{ij}-t)\right\}^{2}K\left(\frac{T_{ij}-t}{h_i}\right),\label{eq:LLE-X}
	\end{equation}
	where $K$ is a compactly supported symmetric density function and $h_i$ is the bandwidth. Calculation shows that  $\hat{b}_{1}=(R_{0}S_{2}-R_{1}S_{1})/(S_{0}S_{2}-S_{1}^{2})$, 
	where for $r=0,1$ and $2$,
	\[
	S_{r}(t)=\frac{1}{m_ih_i}\sum_{j=1}^{m_{i}}K\left(\frac{T_{ij}-t}{h_i}\right)\left(\frac{T_{ij}-t}{h_{i}}\right)^{r}, R_{r}(t)=\frac{1}{m_ih_i}\sum_{j=1}^{m_{i}}K\left(\frac{T_{ij}-t}{h_i}\right)\left(\frac{T_{ij}-t}{h_{i}}\right)^{r}X_{ij}^{\ast}.
	\]
	
	The estimate $\hat{b}_1$ does not have a finite mean squared error, as its denominator 
	is zero with a positive probability for a finite sample. To overcome this issue, we adopt the technique of ridging \citep{Fan1993,Seifert1996,Hall1997} to estimate $X_i(t)$ by the following ridged local linear estimate  
	\begin{equation}
	\hat{X}_{i}(t)=\frac{R_{0}S_{2}-R_{1}S_{1}}{S_{0}S_{2}-S_{1}^{2}+\delta 1_{\{|S_{0}S_{2}-S_{1}^{2}|<\delta\}}},\label{eq:LLE-Xt}
	\end{equation}
	where $\delta>0$ is a sufficiently small constant that depends on $m_i$, e.g., $\delta={m}_i^{-2}$.
	
	When $\sup_i m_i$ is relatively small or bounded by a constant, a scenario commonly referred to as the sparse design, to recover $X_i$, the procedure proposed by \cite{Yao2005a} can be adopted to recover individual functions. We refer readers to Supplementary Material  for the details of such procedure. 
	
	\subsection{Step II: Estimation of the Manifold Dimension and Tangent Space}
	To characterize the manifold structure, we shall first estimate the intrinsic
	dimension $d$ of the manifold $\manifold$. We adopt the maximum
	likelihood estimator proposed by \citet{Levina2004}, substituting the unobservable $X_{i}$ with the contaminated version $\hat{X}_{i}$. For a given $x\in\manifold$, define $\hat{G}_{i}(x)=\|x-\hat{X}_{i}\|_{\ltwo}$ and let $\hat{G}_{(k)}(x)$
	be the $k$th order statistic of $\hat{G}_{1}(x),\ldots,\hat{G}_{n}(x)$. Then the intrinsic dimension $d$ is estimated by 
	\begin{equation}\label{eq:mle-dim-0}
	\hat{d}=\frac{1}{k_{2}-k_{1}+1}\sum_{k=k_{1}}^{k_{2}}\hat{d}_{k},
	\end{equation}
	with 
	\begin{equation}
	\hat{d}_{k}=\frac{1}{n}\sum_{i=1}^{n}\hat{d}_{k}(\hat{X}_{i}), \qquad  \hat{d}_{k}(x)=\left\{\frac{1}{k-1}\sum_{j=1}^{k-1}\log\frac{\hat{G}_{(k)}(x)+\Delta}{\hat{G}_{(j)}(x)+\Delta}\right\}^{-1},\label{eq:mle-dim}
	\end{equation}
	where $\Delta$ is a positive constant depending on $n$, and $k_{1},k_2$ are tuning parameters. This $\Delta$ regularizes  $\hat{d}_k(x)$ in order to overcome the additional variability introduced by the contamination on the predictor.  We conveniently set $\Delta=1/\log \bar m$ with $\bar m=n^{-1}\sum_{i=1}^n m_i$, while refer readers
	to \citet{Levina2004} for the choice of $k_{1}$ and $k_{2}$. {When the observed data are sparsely sampled, the distance $\hat{G}_i(x)$ can be better estimated by the procedure of \cite{Peng2008}.}
	
	Now we proceed to estimate the
	tangent space at the given point $x$ as follows.
	\begin{itemize}
		\item A neighborhood of $x$ is determined by a tuning parameter $\hpca>0$,
		denoted by $\estneighbor[\hpca]=\{\hat{X}_{i}:\|x-\hat{X}_{i}\|_{\ltwo}<\hpca,\; i=1,\ldots,n\}$. 
		\item Compute the local empirical covariance function 
		\begin{eqnarray} \label{eq:localcov}
		\hat{\covarop}_{x}(s,t)=\frac{1}{|\estneighbor[\hpca]|}\sum_{\hat{X}\in\estneighbor[\hpca]}\{\hat{X}(s)-\hat{\mu}_{x}(s)\}\{\hat{X}(t)-\hat{\mu}_{x}(t)\}
		\end{eqnarray}
		 and obtain the eigenfunctions $\hat{\varphi}_{1},\hat{\varphi}_{2},\ldots,\hat{\varphi}_{\hat{d}}$ corresponding to the first $\hat{d}$ leading eigenvalues, where $\hat{\mu}_x=|\estneighbor[\hpca]|^{-1} \sum_{\hat{X}\in\estneighbor[\hpca]} \hat{X}$ is the local mean function and $|\estneighbor[\hpca]|$ denotes the number of observations in $\estneighbor[\hpca]$. 
		\item Estimate the tangent space at $x$ by $\widehat{\txm}=\mathrm{span}\{\hat{\varphi}_{1},\hat{\varphi}_{2},\ldots,\hat{\varphi}_{\hat{d}}\}$,
		the linear space spanned by the first $\hat{d}$ estimated eigenfunctions.
	\end{itemize}
	
	\subsection{Step III: Local Linear Regression on the Tangent Space}
	Finally, we utilize the local manifold structure by projecting all $\hat{X}_{i}$ onto the estimated tangent space $\widehat{\txm}$
	and obtain the local coordinate $\hat{\xi}_{i}=(\langle\hat{X}_{i},\hat{\varphi}_{1}\rangle,\ldots,\langle\hat{X}_{i},\hat{\varphi}_{\hat{d}}\rangle)^{T}$ for $\hat{X}_i$.
	Then, the estimate of $g(x)$ is given by 
	\begin{equation}
	\hat{g}(x)={e}_{1}^{T}(\hat{\bX}^{T}\hat{\bW}\hat{\bX})^{-1}\hat{\bX}^{T}\hat{\bW}\mathcal{Y}, \quad\quad
	\hat{\bX}=\begin{pmatrix} 1 & 1 & \cdots & 1\\
	\hat{\bx}_{1} & \hat{\bx}_{2} & \cdots & \hat{\bx}_{n}
	\end{pmatrix}^T, 
	\label{eq:estimate-gx}
	\end{equation}
	$\hat{\bW}=\mathrm{diag}(K_{\hreg}(\|x-\hat{X}_{1}\|_{\hilbert}),K_{\hreg}(\|x-\hat{X}_{2}\|_{\hilbert}),\ldots,K_{\hreg}(\|x-\hat{X}_{n}\|_{\hilbert}))$
	with $K_{h}(t)=K(t/h)/h^{\hat{d}}$  and the bandwidth $\hreg$, $\mathcal{Y}=(Y_{1},\ldots,Y_{n})^T$,
	and ${e}_{1}^{T}=(1,0,\ldots,0)$ is an $n\times1$ vector. {Here, the matrix $\hat{Q}$ incorporates the estimated geometric structure that is encoded by  the local eigenbasis $\hat{\varphi}_1,\ldots,\hat{\varphi}_{\hat{d}}$.}
	We emphasize that, in the above estimation procedure \textcolor{black}{which is illustrated by the diagram in the left panel of Figure \ref{fig:illustration}}, all steps are
	based on the contaminated sample $\{\hat{X}_1, \ldots, \hat{X}_n\}$, rather than the
	unavailable functions $X_{1},\ldots,X_n$. When the predictor $x$ is also only measured at $m_x$ discrete points $t_1,\ldots,t_{m_x}$, we impute it by the procedures in Section \ref{subsec:stepI}, and replace $x$ in \eqref{eq:mle-dim}--\eqref{eq:estimate-gx} with the imputed curve $\tilde{x}$  to obtain an estimate of $g(\tilde{x})$.

	\subsection{Tuning Parameter Selection}
	There are several tuning parameters to be determined in our estimation procedure. For the parameters $k_1$ and $k_2$ in \eqref{eq:mle-dim-0} to estimate the intrinsic dimension, $k_1=10$ and $k_2=20$ are suggested by \cite{Levina2004}. However, we found that $k_1=20$ and $k_2=30$ work better generally in our setting, perhaps partially due to the contamination that requires a relatively larger local neighborhood to offset.
	
    For the individual smoothing presented in Section \ref{subsec:stepI}, we adopt the following leave-one-out cross-validation to select the bandwidth $h_i$	\citep{Fan1996,Lee1999}. Let $\hat{X}_{i,h,-j}(x)$ be the leave-one-out estimate of $X_i(t)$, i.e., the estimate computed according to \eqref{eq:LLE-Xt} using all of $(T_{i1},X_{i1}^\ast),\ldots,(T_{im_i},X_{im_i}^\ast)$ but $(T_{ij},X_{ij}^\ast)$. We then select $h_i$ from a pool of candidates to  minimize the cross-validation error $\textsc{cv}(h)=\sum_{j=1}^{m_i}\{X_{ij}^\ast-\hat{X}_{i,h,-j}(T_{ij})\}^2.$

	For the bandwidths $h_{pca}$ in \eqref{eq:localcov} and $h_{reg}$ in \eqref{eq:estimate-gx}, we choose the pair $(h_{pca},h_{reg})$  from a pool $\mathbb{H}$ of candidate pairs to minimize the following leave-one-out cross-validation error
	 $\textsc{cv}(h_{pca},h_{reg})=\sum_{i=1}^n\{Y_i-\hat{g}_{h_{pca},h_{reg},-i}(\hat{X}_i)\}^2,$ 
where 
 $\hat{g}_{h_{pca},h_{reg},-i}$ denotes the leave-one-out estimate of $g$ with parameters $h_{pca},h_{reg}$ without using the pair $(\hat{X}_i,Y_i)$. The pool $\mathbb{H}$ shall be constructed in the way that  every $\hat{\mathcal{N}}_{\ltwo}(h_{pca},\hat{X}_i)$ contains at least $\hat{d}+1$ samples for every pair $(h_{pca},h_{reg})$ in $\mathbb{H}$ to ensure sufficient data for local estimation.

 \begin{figure}
 	\begin{center}
 		\resizebox{2.5in}{1.2in}{
 			\input{Algorithm}
 		}
 		\hspace{0.3in}
 		\resizebox{2.5in}{1.6in}{
 			\begin{tikzpicture}
\pgftransformxscale{1.000000}
\pgftransformyscale{-1.000000}
\definecolor{dialinecolor}{rgb}{0.000000, 0.000000, 0.000000}
\pgfsetstrokecolor{dialinecolor}
\definecolor{dialinecolor}{rgb}{1.000000, 1.000000, 1.000000}
\pgfsetfillcolor{dialinecolor}
\pgfsetlinewidth{0.100000\du}
\pgfsetdash{}{0pt}
\pgfsetdash{}{0pt}
\pgfsetbuttcap
{
\definecolor{dialinecolor}{rgb}{0.000000, 0.000000, 0.000000}
\pgfsetfillcolor{dialinecolor}
\definecolor{dialinecolor}{rgb}{0.000000, 0.000000, 0.000000}
\pgfsetstrokecolor{dialinecolor}
\draw (-4.975000\du,11.475000\du)--(-11.012500\du,17.375000\du);
}
\pgfsetlinewidth{0.100000\du}
\pgfsetdash{}{0pt}
\pgfsetdash{}{0pt}
\pgfsetbuttcap
{
\definecolor{dialinecolor}{rgb}{0.000000, 0.000000, 0.000000}
\pgfsetfillcolor{dialinecolor}
\definecolor{dialinecolor}{rgb}{0.000000, 0.000000, 0.000000}
\pgfsetstrokecolor{dialinecolor}
\draw (-4.984295\du,11.520705\du)--(14.757481\du,11.500000\du);
}
\pgfsetlinewidth{0.100000\du}
\pgfsetdash{}{0pt}
\pgfsetdash{}{0pt}
\pgfsetbuttcap
{
\definecolor{dialinecolor}{rgb}{0,0,0}
\pgfsetfillcolor{dialinecolor}
\pgfsetarrowsend{to}
\definecolor{dialinecolor}{rgb}{0,0,0}
\pgfsetstrokecolor{dialinecolor}
\draw (-2.275000\du,13.300000\du)--(9.157481\du,13.300000\du);
}
\pgfsetlinewidth{0.100000\du}
\pgfsetdash{}{0pt}
\pgfsetdash{}{0pt}
\pgfsetbuttcap
{
\definecolor{dialinecolor}{rgb}{0,0,0}
\pgfsetfillcolor{dialinecolor}
\pgfsetarrowsend{to}
\definecolor{dialinecolor}{rgb}{0,0,0}
\pgfsetstrokecolor{dialinecolor}
\draw (-2.379685\du,13.290668\du)--(-5.562500\du,16.925000\du);
}
\pgfsetlinewidth{0.100000\du}
\pgfsetdash{}{0pt}
\pgfsetdash{}{0pt}
\pgfsetbuttcap
{
\definecolor{dialinecolor}{rgb}{0,0,0}
\pgfsetfillcolor{dialinecolor}
\definecolor{dialinecolor}{rgb}{0,0,0}
\pgfsetstrokecolor{dialinecolor}
\draw (-2.300000\du,14.150000\du)--(-1.600000\du,13.250000\du);
}
\pgfsetlinewidth{0.100000\du}
\pgfsetdash{}{0pt}
\pgfsetdash{}{0pt}
\pgfsetbuttcap
{
\definecolor{dialinecolor}{rgb}{0,0,0}
\pgfsetfillcolor{dialinecolor}
\definecolor{dialinecolor}{rgb}{0,0,0}
\pgfsetstrokecolor{dialinecolor}
\draw (-2.267519\du,14.125000\du)--(-3.050310\du,14.120072\du);
}
\pgfsetlinewidth{0.100000\du}
\pgfsetdash{}{0pt}
\pgfsetdash{}{0pt}
\pgfsetbuttcap
{
\definecolor{dialinecolor}{rgb}{0.000000, 0.000000, 1.000000}
\pgfsetfillcolor{dialinecolor}
\pgfsetarrowsend{to}
\definecolor{dialinecolor}{rgb}{0.000000, 0.000000, 0.000000}
\pgfsetstrokecolor{dialinecolor}
\draw (-2.387500\du,13.300000\du)--(4.662500\du,8.200000\du);
}
\pgfsetlinewidth{0.100000\du}
\pgfsetdash{}{0pt}
\pgfsetdash{}{0pt}
\pgfsetbuttcap
{
\definecolor{dialinecolor}{rgb}{0.000000, 0.000000, 0.000000}
\pgfsetfillcolor{dialinecolor}
\definecolor{dialinecolor}{rgb}{0.000000, 0.000000, 0.000000}
\pgfsetstrokecolor{dialinecolor}
\draw (-10.917519\du,17.325000\du)--(8.882481\du,17.350000\du);
}
\pgfsetlinewidth{0.100000\du}
\pgfsetdash{}{0pt}
\pgfsetdash{}{0pt}
\pgfsetbuttcap
{
\definecolor{dialinecolor}{rgb}{0.000000, 0.000000, 0.000000}
\pgfsetfillcolor{dialinecolor}
\definecolor{dialinecolor}{rgb}{0.000000, 0.000000, 0.000000}
\pgfsetstrokecolor{dialinecolor}
\draw (14.803205\du,11.470705\du)--(8.787500\du,17.350000\du);
}
\pgfsetlinewidth{0.100000\du}
\pgfsetdash{{\pgflinewidth}{0.200000\du}}{0cm}
\pgfsetdash{{\pgflinewidth}{0.200000\du}}{0cm}
\pgfsetbuttcap
{
\definecolor{dialinecolor}{rgb}{0.000000, 0.000000, 0.000000}
\pgfsetfillcolor{dialinecolor}
\definecolor{dialinecolor}{rgb}{0.000000, 0.000000, 0.000000}
\pgfsetstrokecolor{dialinecolor}
\draw (4.607481\du,8.325000\du)--(4.612500\du,15.300000\du);
}
\pgfsetlinewidth{0.100000\du}
\pgfsetdash{{\pgflinewidth}{0.200000\du}}{0cm}
\pgfsetdash{{\pgflinewidth}{0.200000\du}}{0cm}
\pgfsetbuttcap
{
\definecolor{dialinecolor}{rgb}{0.000000, 0.000000, 0.000000}
\pgfsetfillcolor{dialinecolor}
\definecolor{dialinecolor}{rgb}{0.000000, 0.000000, 0.000000}
\pgfsetstrokecolor{dialinecolor}
\draw (-2.337500\du,13.300000\du)--(4.662500\du,15.200000\du);
}
\pgfsetlinewidth{0.100000\du}
\pgfsetdash{{\pgflinewidth}{0.200000\du}}{0cm}
\pgfsetdash{{\pgflinewidth}{0.200000\du}}{0cm}
\pgfsetbuttcap
{
\definecolor{dialinecolor}{rgb}{0.000000, 0.000000, 0.000000}
\pgfsetfillcolor{dialinecolor}
\definecolor{dialinecolor}{rgb}{0.000000, 0.000000, 0.000000}
\pgfsetstrokecolor{dialinecolor}
\draw (8.612500\du,13.400000\du)--(4.612500\du,15.250000\du);
}
\definecolor{dialinecolor}{rgb}{0.000000, 0.000000, 0.000000}
\pgfsetstrokecolor{dialinecolor}
\node[anchor=west] at (4.562500\du,10.150000\du){};
\definecolor{dialinecolor}{rgb}{0.000000, 0.000000, 1.000000}
\pgfsetstrokecolor{dialinecolor}
\node[anchor=west,rotate=30] at (-0.012500\du,9.500000\du){{$\hat{\varphi}_1$}};
\definecolor{dialinecolor}{rgb}{0.000000, 0.000000, 0.000000}
\pgfsetstrokecolor{dialinecolor}
\node[anchor=west] at (6.612500\du,15.300000\du){};
\definecolor{dialinecolor}{rgb}{0,0,0}
\pgfsetstrokecolor{dialinecolor}
\node[anchor=west] at (6.212500\du,12.500000\du){{$\iota_\ast\phi_1$}};
\definecolor{dialinecolor}{rgb}{0,0,0}
\pgfsetstrokecolor{dialinecolor}
\node[anchor=west,rotate=60] at (-5.587500\du,15.700000\du){{$\iota_\ast\phi_2$}};
\definecolor{dialinecolor}{rgb}{0.000000, 0.000000, 0.000000}
\pgfsetstrokecolor{dialinecolor}
\node[anchor=west] at (-6.812500\du,18.350000\du){tangent space {$T_x\mathcal{M}$}};
\pgfsetlinewidth{0.100000\du}
\pgfsetdash{}{0pt}
\pgfsetdash{}{0pt}
\pgfsetbuttcap
{
\definecolor{dialinecolor}{rgb}{0.000000, 0.000000, 0.000000}
\pgfsetfillcolor{dialinecolor}
\definecolor{dialinecolor}{rgb}{0.000000, 0.000000, 0.000000}
\pgfsetstrokecolor{dialinecolor}
\draw (4.612500\du,14.600000\du)--(5.462500\du,14.100000\du);
}
\pgfsetlinewidth{0.100000\du}
\pgfsetdash{}{0pt}
\pgfsetdash{}{0pt}
\pgfsetbuttcap
{
\definecolor{dialinecolor}{rgb}{0.000000, 0.000000, 0.000000}
\pgfsetfillcolor{dialinecolor}
\definecolor{dialinecolor}{rgb}{0.000000, 0.000000, 0.000000}
\pgfsetstrokecolor{dialinecolor}
\draw (5.401527\du,14.119027\du)--(5.382481\du,14.900000\du);
}
\node[anchor=west,rotate=90] at (5.762500\du,11.550000\du){{\footnotesize $O_P(h_{pca})\mathbf{u}^\perp$}};
\node[anchor=west,rotate=25] at (4.712500\du,16.500000\du){{\scriptsize $O_P(h_{pca}^{3/2})\mathbf{u}$}};
            \path (3.562500\du,11.150000\du) coordinate (u1)
                  (5.312500\du,15.100000\du) coordinate (u2);
                  
\end{tikzpicture}
 		}
 	\end{center}		
 	\caption{\label{fig:illustration} Illustration of functional regression on a manifold (left panel) and illustration of the asymptotic result of \eqref{eq:rate-of-tangent-vector} for $d=2$ (right panel).
 	}
 \end{figure}

	\section{Theoretical Properties \label{sec:Theory}}
	{We focus on the scenario that $\inf_{i}m_i$ increases with the sample size $n$, while leave the one that $\sup_{i}m_i\leq m_0<\infty$ for future research due to elevated challenges.} 
	Without loss of generality, assume $m_{i}\asymp m$ 
	where  $a_n \asymp b_n$ denotes $0 < \lim\inf a_n/b_n<\lim\sup a_n/b_n<\infty$. {We further assume that $\zeta_{ij}$, and similarly, $T_{ij}$ and $X_i$, are independently and identically distributed, while emphasize that the development below can be modified to accommodate fixed designs,  weak dependence and/or heterogeneous distributions. This generality will require considerably heavier technicalities without adding further insight, and is not pursued here.} 
	
	The discrepancy
	between $\hat{X}_{i}$ and $X_{i}$, quantified by $\|\hat{X}_{i}-X_{i}\|_{\ltwo}$,
	is termed the contamination of $X_{i}$. The decay of this contamination is intimately 
	linked to the consistency of our estimates of the intrinsic dimension,
	the tangent space, and eventually the regression functional $g(x)$. Moreover, the convergence rate of $\hat{g}(x)$ is found to exhibit a phase transition phenomenon depending on the interplay between the intrinsic dimension and the decay of contamination. To set the stage, we start with a property of contamination in recovery of  functional data by the individual smoothing approach in Section \ref{subsec:stepI}. Specifically, we study the $p$th moment of contamination when $\hat{X}_{i}$ is the ridged local linear estimate in (\ref{eq:LLE-Xt}).   Our result below for an arbitrary $p$th moment is not present in the literature \citep[e.g.,][for $p=2$ only]{Fan1993}.
	
	Let $\Sigma(\nu,L)$ denote the H\"{o}lder class with an exponent $\nu$
	and an H\"{o}lder constant $L$, which represents the set of $\ell=\lfloor\nu\rfloor$
	times differentiable functions $F$ whose derivative $F^{(\ell)}$
	satisfies $|F^{(\ell)}(t)-F^{(\ell)}(s)|\leq L |t-s|^{\nu-\ell}$ for $s,t\in D,$
	where $\lfloor\nu\rfloor$ denotes the largest integer strictly smaller
	than $\nu$. We require the following mild assumptions, and assume $h_i \asymp h_0$ without loss of generality. 
	\begin{lyxlist}{00.00.00}
		\item [{(A1)}] $K$ is a differentiable kernel with a bounded
		derivative, $\int_{-1}^{1}K(u)du=1$,  $\int_{-1}^{1}uK(u)du=0$, and
		$\int_{-1}^{1}|u|^{p}K(u)du<\infty$ for all $p>0$.
		\item [{(A2)}] The sampling density $f_{T}$ is bounded away from zero
		and infinity, i.e., for some constants $C_{T,1},C_{T,2}\in(0,\infty)$,
		$C_{T,1}=\inf_{t\in D}f_{T}(t)\leq\sup_{t\in D}f_{T}(t)=C_{T,2}$.
		\item [{(A3)}] $X\in\Sigma(\nu,L_{X})$, where $L_X>0$ is a random quantity and 
		the constant $\nu\in(0,2]$ quantifies the smoothness of the process. 
		\item [{(A4)}] For all $r\geq1$, $\expect\sup_{t}|X(t)|^{r}<\infty$,
		$\expect(L_{X})^{r}<\infty$ and $\expect|\zeta|^{r}<\infty$.
	\end{lyxlist}
	The condition  $\expect\sup_{t}|X(t)|^{r}<\infty$
	holds rather generally \citep{Li2010, Zhang2016}, compared to a stronger assumption on $X$ given in (A.1) of 
	\citet{Hall2006a}. The following proposition is an immediate consequence of Lemma S.1 in the Supplementary Material, and hence its proof is omitted.
	\begin{proposition}
		\label{prop:fda} For any $p\ge 1$, assume $\expect|\zeta|^{p}<\infty$.
		Under the assumptions \textup{(A1)--(A3)}, for the estimate $\hat{X}$ in \textup{\eqref{eq:LLE-Xt}} with $h_{0}\asymp m^{-\frac{1}{2\nu+1}}$ and  $\delta=m^{-2}$, we have
		\begin{equation}
		\{\expect(\|\hat{X}-X\|_{\ltwo}^{p}\mid X)\}^{1/p}=O(m^{-\frac{\nu}{2\nu+1}})\left\{\sup_{t}|X(t)|+L_{X}\right\}.\label{eq:fda-2}
		\end{equation}
		Furthermore, if  the assumption (A4) also holds, then 
		$\left(\expect\|\hat{X}-X\|_{\ltwo}^{p}\right)^{1/p}=O(m^{-\frac{\nu}{2\nu+1}}).$
	\end{proposition}
	When $X$ is deterministic as in nonparametric regression, the rate in (\ref{eq:fda-2}) for $p=2$ coincides with that in \citet{Tsybakov2008}. In addition, the $p$th order of the contamination $\|\hat{X}_{i}-X_{i}\|_{\ltwo}$
	decays at a polynomial rate that depends on $\nu$, but not the order $p$.

	To analyze the asymptotic property
	of $\hat{g}(x)$, we make the following assumptions.
	\begin{lyxlist}{00.00.00}
		\item [{(B1)}] The probability density $f$ of $X$ on $\manifold$ satisfies
		$C_{f,1}=\inf_{x\in\mathcal{M}}f(x)\leq\sup_{x\in\mathcal{M}}f(x)=C_{f,2}$
		for some constants $0<C_{f,1}\leq C_{f,2}<\infty$.
		\item [{(B2)}] The regression functional $g$ has 
		a bounded second derivative.
	\end{lyxlist}
For (B1), since the functional predictor resides on a low-dimensional manifold, the existence of a density can be safely assumed. We also make the following  assumption on the imputed trajectories in Section \ref{subsec:stepI}. 
	\begin{lyxlist}{00.00.00}
		\item [{(B3)}] $\hat{X}_{1},\ldots,\hat{X}_{n}$ are independently
		and identically distributed. For some $\beta\in(0,\infty)$ and  all $p\geq1$, $\{\expect(\|\hat{X}-X\|_{\ltwo}^{p}\mid X)\}^{1/p}\leq C_{p}m^{-\beta}\eta(X)$
		for some constant $C_{p}$ depending only on $p$ and some nonnegative
		function $\eta(X)$ depending only on $X$ such that $\expect\{\eta(X)\}^{p}<\infty$.
	\end{lyxlist}
	Under the assumptions (A1)--(A4), by Proposition \ref{prop:fda}, the imputed functions $\hat{X}_1,\ldots,\hat{X}_n$ by individual smoothing via local linear estimation (\ref{eq:LLE-Xt}) satisfies (B3)
	with $\beta=\nu/(2\nu+1)$. Therefore, (B3) could be replaced with the more concrete assumptions  (A1)--(A4).  
	It can be relaxed to accommodate heterogeneous data distributions and weakly dependent functional data by modifying our proofs. Also, it is possible to accommodate imputed functions that are attained by borrowing information across individuals \citep[e.g.,][]{Yao2005a}, which is beyond our scope here and can be a topic of future research.

	The contamination of the predictor $X$ renders the true neighborhood  $\mathcal{N}_{\mathcal{L}^{2}}(\hpca,x)=\{X_{i}:\|X_{i}-x\|_{\ltwo}<\hpca\}$ inaccessible. However, we can show that the contaminated one  $\estneighbor[\hpca]$ is a good estimate; see Section S.2 and Lemma 8 in Supplementary Material for details. Consequently, the local manifold structure can be consistently estimated in the sense of the following theorem.
	\begin{theorem}
		\label{lem:bound-tangent-estimation} Suppose that the assumptions \textup{(B1)} and \textup{(B3)} hold.   
		\begin{enumerate}[label=\textup{(\alph*)}]
			\item  $\hat{d}$ is a consistent estimator of $d$ when $\min\{k_1,k_2\}\rightarrow\infty$ and $\max\{k_1,k_2\}/m\rightarrow0$. \label{thm:a}
			\item\label{thm:b} If $h_{pca}\rightarrow 0$ and $h_{pca} \asympgt \max\{m^{-\beta+\epsilon},n^{-1/(d+2)}\}$ for an arbitrarily small but fixed constant $\epsilon>0$, then
			the eigenbasis $\{\hat{\varphi}_{k}\}_{k=1}^{d}$ derived from $\hat{\mathcal{C}}_x$ in \textup{(\ref{eq:localcov})} is close 
			to an orthonormal basis $\{\phi_{k}\}_{k=1}^{d}$ of $\txm$, in the sense that, for each 
			$x\in\mathcal{M}$,
			\begin{equation}
			\hat{\varphi}_{k}=\phi_{k}+\Op(\hpcad{3/2}){u}_{k}+\Op(\hpca){u}_{k}^{\perp}\quad(k=1,\ldots,d),\label{eq:rate-of-tangent-vector}
			\end{equation}
			where  ${u}_{k}\in\txm$,
			${u}_{k}^{\perp}\perp\txm$, and $\|{u}_{k}\|_{\ltwo}=\|{u}_{k}^{\perp}\|_{\ltwo}=1$.
		\end{enumerate}
	\end{theorem}
	In light of Theorem \ref{lem:bound-tangent-estimation}\ref{thm:a}, we shall from now on present the subsequent results by conditioning on the event
	$\hat{d}=d$. For part (b) that is  illustrated in the right panel of Figure \ref{fig:illustration}, the condition  $h_{pca} \asympgt m^{-\beta+\epsilon}$ suggests that $h_{pca}$ shall be larger than the contamination by an arbitrarily small polynomial order of $m$. This is required to ensure that the discrepancy between the estimated local neighborhood $\hat{N}_{\ltwo}(h_{pca},x)$ and the uncontaminated neighborhood $\mathcal{N}_{\ltwo}(h_{pca},x)=\{X_i:\|x-X_i\|_{\ltwo}<h_{pca},i=1,\ldots,n\}$ is asymptotically negligible, suggested by Lemma 8 in  Supplementary Material. The curvature at $x$ is a constant that is absorbed into the $\Op$ terms, and thus 
	does not influence the asymptotic rate. However, practically it is often more difficult to estimate the tangent structure at a point with larger curvature.}
	
	We are ready to state the results on the estimated regression functional. Recall
	that $\hat{g}(x)$ in (\ref{eq:estimate-gx}) is obtained by applying
	the local linear smoother to the coordinates of contaminated predictors
	within the estimated tangent space at $x$. It is well known that
	the local linear estimator does not suffer from boundary effects, i.e.,
	the first order behavior of the estimator on the boundary
	is the same as in the interior \citep{Fan1992}. However, the contamination
	of the predictor has different impact, and we shall address the interior  and boundary cases separately. Denote $\observed=\{(X_{1},\hat{X}_{1}),\ldots,(X_{n},\hat{X}_{n})\}$ and $\manifold_{h}=\{x\in\manifold:\inf_{y\in\partial\manifold}\mathfrak{d}(x,y)\leq h\}$,
	where $\partial\manifold$ denotes the boundary of $\manifold$ and \textcolor{black}{$\mathfrak{d}(\cdot, \cdot)$ denotes the distance function on $\manifold$}. For
	points sufficiently far away from the boundary of $\manifold$, we
	have the following result about the convergence rate of the estimator $\hat{g}(x)$. 
	
	\begin{theorem}
		\label{thm:MSE-interior}Assume that \textup{(A1)} and \textup{(B1)--(B3)} hold. Let $x\in\manifold\backslash\manifold_{\hreg}$
		and $\hpca$ satisfies the conditions of Theorem \ref{lem:bound-tangent-estimation}\textup{\ref{thm:b}}.
		For an arbitrarily small but fixed constant $\epsilon>0$, suppose that $h_{reg}\rightarrow 0$, $h_{reg}>h_{pca}$, and  $\min\{nh_{reg},m^{\beta}h^{5/3+\epsilon }_{reg}\} \rightarrow\infty$. Then 
		\begin{equation}\label{eq:rate-gx-0}
		\expect\left[ \left\{\hat{g}(x)-g(x)\right\}^{2}\mid\observed\right] =\Op\left(h^4+\frac{1}{m^{2\beta}h_{reg}^{2+2\epsilon }}+\frac{1}{nh^{d}}\right).
		\end{equation}
		In addition, if $h_{pca} \asymp \max\{m^{-\beta},n^{-1/(d+2)}\}$, and $h_{reg}\asymp n^{-1/(d+4)}$ when $m \asympgt n^{(3+\epsilon )/\{\beta(d+4)\}}$ and $h_{reg}\asymp m^{-\beta/(3+\epsilon)}$ otherwise, then 
		\begin{equation}
		\expect\left[ \left\{\hat{g}(x)-g(x)\right\}^{2}\mid\observed\right] =\Op\left(n^{-\frac{4}{d+4}}+m^{-\frac{4\beta}{3+\epsilon}}\right).\label{eq:rate-gx}
		\end{equation}
	\end{theorem}
	We emphasize the following observations from this theorem. First, according to our analysis in Supplementary Material, the first two terms on the right hand side of \eqref{eq:rate-gx-0} correspond to the bias while the last term stems from the variability of the estimator. This suggests that, under the conditions of the theorem, the contamination has  impact on the asymptotic bias but not the variance. Second, the convergence rate of $\hat{g}(x)$
	is a polynomial of the sample size $n$ and the sampling rate $m$. This is in contrast with traditional functional nonparametric regression methods that do not exploit the intrinsic structure and thus cannot reach a polynomial rate of convergence. 
	
	Third, the rate in \eqref{eq:rate-gx} consists of two terms, one related to the intrinsic dimension $d$ and the sample size $n$, and the other related to  $m$ and $\beta$ that together characterize the contamination of the predictor.  As $\epsilon>0$ is arbitrary, the transition of these two terms occurs at the rate $\mathfrak{m}_0\asymp n^{3/\{\beta(d+4)\}}$. When the sampling rate falls below $\mathfrak{m}_0$, the contamination term dominates the convergence rate in  \eqref{eq:rate-gx}. Otherwise, the intrinsic dimension and sample size determine the rate. This phase transition, although sharing the similar spirit of \cite{Cai2011} and \cite{Zhang2016}, has a different interpretation, as follows. When the contamination level is low, the manifold structure can be estimated reliably and utilized for regression. In contrast, when the contamination is in a high level, for example, where $m$ or $\beta$ is small, the manifold structure is buried by noise and cannot be well exploited. 
	Finally, it is observed that the phase transition
	threshold $m_{0}$ increases with the intrinsic dimension $d$ that indicates the complexity of a manifold. This interesting finding suggests that, although a complex manifold makes the estimation more challenging, for example, leading to a slower rate, such manifold is more resistant to contamination.

	In our setup, the actual observed predictor is ${\mathbb{X}}_{i}=\{(T_{i1},X_{i1}^{\ast}),\ldots,(T_{im},X_{im_i}^{\ast})\}$ and is an $m_i$-dimensional random vector. Moreover, the distribution of this random vector is {fully} supported on $\real^{m_i}$  due to the presence of the noise $\zeta_{ij}$, and thus  the support of the distribution of the recovered trajectory $\hat{X}_{i}$ is also $m_i$-dimensional.  Smoothness of functional data could help tighten the distribution of $\hat{X}_{i}$, but does not reduce its dimension. As  $m_i$ goes to infinity, it might then raise a serious concern of the curse of dimensionality. In this sense, the polynomial rate and phase transition phenomenon in Theorem \ref{thm:MSE-interior} are remarkable: when $\inf_i m_i$ surpasses certain threshold, by exploiting the low-dimensional manifold structure, the growing dimension of the contamination can be defeated with the aid of smoothness.
	
	The following theorem characterizes the behavior of $\hat{g}$ on the boundary of $\manifold$.
	\begin{theorem}
		\label{thm:MSE-boundary}
		Assume that \textup{(A1)} and \textup{(B1)--(B3)} hold. Let $x\in\manifold_{\hreg}$ 
		and $\hpca$ satisfies the conditions of Theorem \ref{lem:bound-tangent-estimation}\textup{\ref{thm:b}}.
		For an arbitrarily small but fixed constant  $\epsilon>0$, suppose that $h_{reg}\rightarrow 0$, $h_{reg}>h_{pca}$, and  $\min\{nh_{reg},m^{\beta}h^{3+\epsilon }_{reg}\} \rightarrow\infty$. Then 
		\begin{equation*}
		\expect\left[ \left\{\hat{g}(x)-g(x)\right\}^{2}\mid\observed\right] =\Op\left(h^4+\frac{1}{m^{2\beta}h_{reg}^{4+2\epsilon }}+\frac{1}{nh^{d}}\right).
		\end{equation*}
		In addition, if $h_{pca} \asymp \max\{m^{-\beta},n^{-1/(d+2)}\}$, and $h_{reg}\asymp n^{-1/(d+4)}$ when $m \asympgt n^{(4+\epsilon )/\{\beta(d+4)\}}$ and $h_{reg}\asymp m^{-\beta/(4+\epsilon)}$ otherwise, then 
		\begin{equation}
		\expect\left[ \left\{\hat{g}(x)-g(x)\right\}^{2}\mid\observed\right] =\Op\left(n^{-\frac{4}{d+4}}+m^{-\frac{4\beta}{4+\epsilon}}\right).\label{eq:rate-gx-b}
		\end{equation}
	\end{theorem}
	By comparing the above with Theorem \ref{thm:MSE-interior}, we see that the effect of the intrinsic dimension on convergence
	is the same, regardless where $\hat{g}$ is evaluated on the manifold.
	However, the effect of contamination behaves differently, due to the fact that the second order behavior of the local linear estimator depends on the location and needs to be  considered  when there is contamination of
	$X$. Moreover, we see that  the phase transition
	occurs at $\mathfrak m_{1}=n^{4/\{\beta(d+4)\}} \gg \mathfrak m_0$, and when the contamination  dominates, the convergence is slightly slower for boundary points
	than for interior points. This is the price we pay for the boundary effect when the predictor is contaminated, which is in contrast with the classical result on the local linear estimator \citep{Fan1993}.

	\section{Simulation Study\label{sec:Simulation-Studies}}
	
	To demonstrate the performance of our framework,
	we conduct simulation studies for three different
	manifolds, namely, the three-dimensional rotation group $SO(3)$,
	the Klein bottle and the mixture of two Gaussian densities.
	\begin{itemize}
		\setlength\itemsep{0.01em}
		\item {${SO(3)}$ manifold}: we set $X_{i}(t)=\sum_{k=1}^{9}z_{ik}b_{k}(t)$,
		where $b_{2\ell-1}(t)=\cos\{(2\ell-1)\pi t/10\}/{5}^{1/2}$ and 
		$b_{2\ell}(t)=\sin\{(2\ell-1)\pi t/10\}/{5}^{1/2}$. To generate the random
		variables $z_{ik}$, for a vector ${r}=(r_{1},r_{2},r_{3})$ and a variable $\theta\in\real$, we define  
		\[
		{R}({r},\theta)=(1-\cos\theta){r}{r}^{T}+\begin{pmatrix}\cos\theta & -r_{3}\sin\theta & r_{2}\sin\theta\\
		r_{3}\sin\theta & \cos\theta & -r_{1}\sin\theta\\
		-r_{2}\sin\theta & r_{1}\sin\theta & \cos\theta
		\end{pmatrix}.
		\]
		Denoting ${e}_{2}=(0,1,0)^{T}$ and ${e}_{3}=(0,0,1)^{T}$, we set $(z_{i1},\ldots,z_{i9})^T=\mathrm{vec}({Z}_{i})$
		with Euler angle parameterization
		${Z}_{i}={R}({e}_{3},u_{i}){R}({e}_{2},v_{i}){R}({e}_{3},w_{i})$, 
		where $(u_i, v_i)$ are uniformly sampled from  the two-dimensional sphere $S^2=[0,2\pi)\times[0,\pi]$, and $w_i$ are uniformly sampled from the unit circle $S^1=[0,2\pi)$. 
		
		\item {Klein bottle}: we set $X_{i}(t)=\sum_{k=1}^{4}z_{ik}b_{k}(t)$ 
		with $b_{k}(t)$ as in the $SO(3)$ setting. We set $z_{i1}=(2\cos v_{i}+1)\cos u_{i}$,
		$z_{i2}=(2\cos v_{i}+1)\sin u_{i}$, $z_{i3}=2\sin v_{i}\cos(u_{i}/2)$
		and $z_{i4}=2\sin v_{i}\sin(u_{i}/2)$, where $u_{i}$ and $v_{i}$ independently sampled from the uniform distribution on $(0,2\pi)$.
		Here $(u,v)\mapsto(z_{1},z_{2},z_{3},z_{4})$ is
		a parameterization of the Klein bottle with an intrinsic dimension $d=2$. 
		\item {Gaussian mixture}: we set $X_{i}$ to $X_{i}(t)=\exp\{-(t-u_i)^{2}/2\}/{(2\pi)}^{1/2}+\exp\{-(t-v_i)^{2}/2\}/{(2\pi)}^{1/2}$
		with $(v_{1}, v_{2})^T$ uniformly sampled from a circle with diameter
		$0.5$, similar to that used in \citet{Chen2012}.
	\end{itemize}

The functional predictor $X_{i}$ is observed at $m_i$ points $T_{i1},\ldots,T_{im_i}$ in the interval
$[0,1]$ with heteroscedastic measurement errors $\zeta_{ij}\sim N(0,\sigma_{ij}^{2})$, where $\sigma_{ij}$ is determined  by the signal-to-noise ratio  $\mathrm{snr}_{X}=\mathrm{Var}\{X(T_{ij})|T_{ij}\}/\sigma_{ij}^{2}=4$. The response is
generated by {$Y_{i}=4\sin(4Z_i)\cos(Z_i^2)+2\Gamma(1+Z_i/2)+\varepsilon_{i}$ with $Z_i=\int_{0}^1 X_i^2(t)t\diffop t$ and $\Gamma(\alpha)=\int_0^\infty s^{\alpha-1}e^{-s}\diffop s$.} The noise $\varepsilon_{i}$ added to the response $Y$ is a centered Gaussian variable
with variance $\sigma_{\varepsilon}^{2}$ that is determined by 
the signal-to-noise ratio $\mathrm{snr}_{Y}=\mathrm{var}(Y)/\sigma_{\varepsilon}^{2}=2$. 
	{To see the impact of the manifold structure on regression, we normalize the functional predictor in all settings to the unit scale, i.e., multiplying $X$ by the constant $c=1/(\expect\|X\|^2)^{1/2}$ so that the resultant $X$ satisfies $\expect\|X\|^2=1$. Such scaling does not change the geometric structure of manifolds but the size.}
	In order to account for at least 95\% of variance of data, we find empirically that more than 10 principal components are needed in all settings, i.e., the dimensions of the contaminated data are considerably larger than their intrinsic dimensions.
	
	For evaluation, we
	generate independent test data of size 5000, and compute the root mean square error using the test data. {In the test data, each predictor is also discretely measured and contaminated by noise in the same way of the training sample.}
	We compare our method with nonparametric estimators based on functional Nadaraya--Watson smoothing, functional conditional expectation, functional mode, functional conditional median and  multi-method  that averages estimates from the methods of functional conditional expectation, functional mode and functional conditional median \citep{Ferraty2006}. 
	Functional linear regression is also included
	to illustrate the impact of nonlinear relationship. The tuning parameters in these methods, such as the number of
	principal components for functional linear regression and the bandwidth for the nonparametric methods, are selected by 10-fold cross-validation. 
	
	We consider the scenario of dense functional data here, while refer readers to Supplementary Material for simulation studies for sparsely observed data. Specifically, we set $m_i=m=100$ and $T_{ij}=t_j$,  where $t_1,\ldots,t_m$ are  equally spaced over $[0,1]$. Three sample sizes are considered, namely,  $n=250,500,1000$. We repeat each	study 100 times independently, and the results are presented in Table
	\ref{tab:simulation}. {First, we observe that the proposed method enjoys favorable numerical performance in all simulation settings. Second, as the sample size grows, the reduction in root mean square error is more prominent
		for the proposed method than for the others. For example, the relative reduction from $n=250$ ($n=500$, respectively) to $n=500$ ($n=1000$, respectively) is $25.5\%$  ($22.7\%$, respectively)  for our method, but $8.49\%$ ($2.75\%$, respectively) for the functional Nadaraya--Watson estimator. This may provide some numerical evidence that the proposed estimator has a faster convergence rate. Furthermore, it also provides evidence for the polynomial rate stated in Theorem \ref{thm:MSE-interior} and \ref{thm:MSE-boundary}. 
		Based on these theorems the relative reduction is expected to be $1-(n_1/n_2)^{2/(d+4)}$ when the sample size increases from $n_1$ to $n_2$, as the data is sufficiently dense and thus the convergence rate is dominated by the intrinsic dimension. For the setting of Klein bottle, it is about $20.6\%$, and the empirical relative reduction is  $22.7\%$ from $n_1=500$ to $n_2=1000$. Similar observations can be made for other settings. In contrast, the existing kernel methods perform no better than a logarithmic rate, providing numerical evidence for the theory of \citet{Mas2012}.  
		Third, as the intrinsic dimension goes up, the relative reduction in root mean square error for our estimator decreases, suggesting that the intrinsic dimension plays an important role in the convergence rate. Finally, different manifolds result in different constants hidden in the $\Op$ terms in Theorem \ref{thm:MSE-interior} and \ref{thm:MSE-boundary}. For example, those in the $SO(3)$ setting seem relatively smaller than their counterparts in the  setting of Klein bottle according to Table \ref{tab:simulation}.

		\begin{table}
			\def~{\hphantom{0}}
			\tbl{Results of simulation studies for densely observed data}{%
				\resizebox{\textwidth}{!}{\begin{tabular}{cccccccccc}
						\multirow{2}{*}{} & \multicolumn{3}{c}{$SO(3)$ Manifold} & \multicolumn{3}{c}{Klein Bottle} & \multicolumn{3}{c}{Gaussian Mixture}\tabularnewline
					& $n=250$ & $n=500$ & $n=1000$ & $n=250$ & $n=500$ & $n=1000$ & $n=250$ & $n=500$ & $n=1000$\tabularnewline
					FLR & $22.1\, {(0.34)}$ & $21.8\, {(0.23)}$ & $21.6\, {(0.20)}$ & $61.3\, {(0.62)}$ & $61.2\, {(0.39)}$ & $6.09\, {(0.35)}$ & $29.6\, {(1.43)}$ & $29.0\, {(1.26)}$ & $28.8\, {(0.99)}$\tabularnewline
					FNW & $16.2\, {(0.58)}$ & $15.7\, {(0.43)}$ & $15.5\, {(0.32)}$ & $31.8\, {(4.05)}$ & $29.1\, {(1.79)}$ & $28.3\, {(0.68)}$ & $18.7\, {(1.46)}$ & $17.5\, {(0.83)}$ & $17.0\, {(0.65)}$\tabularnewline
					FCE & $15.3\, {(0.66)}$ & $14.1\, {(0.52)}$ & $13.2\, {(0.30)}$ & $29.7\, {(1.46)}$ & $27.1\, {(1.04)}$ & $26.1\, {(0.81)}$ & $21.1\, {(1.32)}$ & $20.4\, {(0.93)}$ & $19.8\, {(0.64)}$\tabularnewline
					FMO & $25.4\, {(1.16)}$ & $23.0\, {(0.94)}$ & $22.0\, {(0.85)}$ & $46.2\, {(3.07)}$ & $41.2\, {(2.20)}$ & $38.3\, {(1.87)}$ & $35.9\, {(2.80)}$ & $33.6\, {(2.05)}$ & $32.2\, {(1.61)}$\tabularnewline
					FCM & $20.2\, {(0.60)}$ & $18.6\, {(0.52)}$ & $17.2\, {(0.35)}$ & $39.1\, {(2.67)}$ & $33.9\, {(1.61)}$ & $30.9\, {(1.02)}$ & $27.3\, {(1.71)}$ & $25.1\, {(1.05)}$ & $23.2\, {(0.83)}$\tabularnewline
					MUL & $18.2\, {(0.59)}$ & $16.6\, {(0.48)}$ & $15.4\, {(0.31)}$ & $34.0\, {(2.13)}$ & $30.0\, {(1.24)}$ & $27.7\, {(0.92)}$ & $24.6\, {(1.49)}$ & $23.1\, {(1.07)}$ & $21.8\, {(0.81)}$\tabularnewline
					{FREM} & $10.1\, {(0.72)}$ & $8.16\, {(0.56)}$ & $6.38\, {(0.25)}$ & $16.5\, {(1.39)}$ & $12.3\, {(1.11)}$ & $9.51\, {(0.74)}$ & $10.5\, {(1.32)}$ & $8.08\, {(0.86)}$ & $6.12\, {(0.75)}$\tabularnewline
		\end{tabular}}}
	\label{tab:simulation}
\begin{tabnote}
{\footnotesize FLR, functional linear regression; FNW, functional Nadaraya--Watson smoothing; FCE, functional conditional expectation; FMO, functional mode, FCM, functional conditional median; MUL,  multi-method; FREM, the proposed functional regression on manifold; MSP, meat spectrometric data; DTI, diffusion tensor imaging data; SBP, systolic blood pressure data.  The numbers outside of parentheses are the Monte Carlo average of root mean square error based on 100 independent simulation replicates, and the numbers in parentheses are the corresponding standard error.}
\end{tabnote}
\end{table}	
	
	\section{Real Data Examples \label{sec:Applications}}
	
	We apply our method to analyze three real datasets. For the
	purpose of evaluation, we train our method on $75\%$ of each dataset
	and reserve the other $25\%$ as test data. The root mean square error is computed on
	the held-out test data. 
	We repeat this 100 times based on random partitions of the 	datasets, and summarize the results in Table \ref{tab:spectrometric}. 
	
	The first application is to predict the fat content of a piece of meat
	based on a spectrometric curve for the meat using the Tecator dataset with 215 meat samples \citep{Ferraty2006}.
	For each sample, the spectrometric
	curve for a piece of finely chopped pure meat was measured at 100
	different wavelengths from 850 to 1050nm. Along with  the spectrometric
	curves, the fat content for each piece of meat was recorded. Comparing to the analytic chemistry required for measuring the fat content, obtaining a spectrometric curve is less time and cost consuming. As in \citet{Ferraty2006}, we predict the fat content based on the first derivative curves approximated by the difference quotient between measurements at adjacent wavelengths, shown in the left panel of Figure \ref{fig:data}. It is
	seen that there are some striking patterns around the middle  
	wavelengths. The proposed method is able to capture these patterns
	by a low-dimensional manifold structure. For example,  functional linear regression uses 15.7  principal components on average with a standard error 1.07, while the intrinsic dimension estimated by our method is 5.05 with a standard error 0.62. Thus, our method predicts the fat content more accurately than the others by a significant margin according to Table \ref{tab:spectrometric}.
	
	\begin{figure}
		\begin{centering}
			\begin{minipage}[t]{0.3\textwidth}%
				\begin{center}
					\includegraphics[scale=0.4]{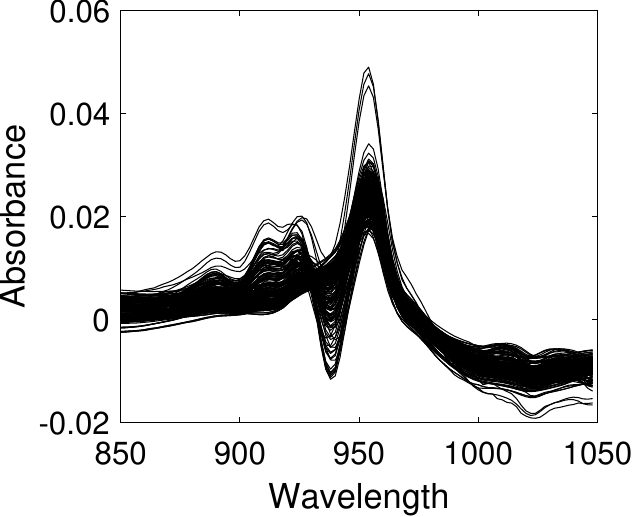}
					\par\end{center}%
			\end{minipage} 
			\begin{minipage}[t]{0.3\textwidth}%
				\begin{center}
					\includegraphics[scale=0.4]{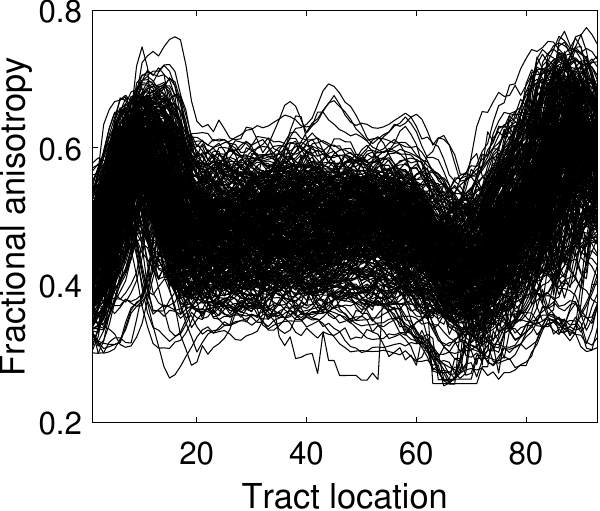}
					\par\end{center}%
			\end{minipage}
			\begin{minipage}[t]{0.3\textwidth} 
				\begin{center}
					\includegraphics[scale=0.4]{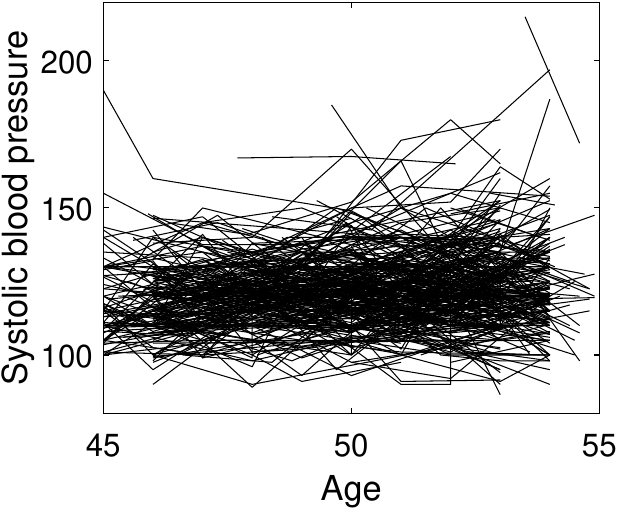}
					\par\end{center}%
			\end{minipage}
			\par\end{centering}
		\protect\caption{ \label{fig:data} first derivatives of meat spectrometric curves (left panel), fractional anisotropy profiles (middle panel), and  
			systolic blood pressure over time (right panel).
		}
	\end{figure}

	{The second example studies the relationship between cognitive function and brain microstructure in the corpus callosum  of patients with multiple sclerosis, a common demyelinating disease caused by inflammation in the brain. Demyelination refers to the damage to myelin that protects axons and helps nerve signal to travel faster. It occurs in the white matter of the brain and can potentially lead to loss of mobility or even cognitive impairment \citep{Jongen2012}. Diffusion tensor imaging, a technique that can produce high-resolution images of white matter tissues by tracing water diffusion within the tissues,  is an important method to examine potential myelin damage in the brain. For example, from such images, some properties of white matter, such as  fractional anisotropy of water diffusion, can be derived. It has been shown that fractional anisotropy is related to multiple sclerosis \citep{Ibrahim2011}.}
	
	{To predict cognitive performance based on fractional anisotropy profiles, we utilize the data collected at Johns Hopkins University and the Kennedy-Krieger Institute. The data contains $n=340$ profiles from multiple sclerosis patients and paced auditory serial addition test  scores that quantify cognitive function \citep{Gronwall1977}, where each profile was recorded at a grid of 93 points. In the middle panel of Figure \ref{fig:data}, we show all fractional anisotropy profiles, and observe that the data is considerably more complex than the spectrometric data. The average of estimated intrinsic dimensions is 5.82 with a standard error 0.098. By contrast, the average number of principal components for functional linear regression is 11.98 with a 
		standard error 5.22. According to Table \ref{tab:spectrometric}, our method enjoys the most accurate prediction, while all other functional nonparametric methods deteriorate substantially. }
	
	Our third example concerns systolic blood pressure  of healthy
	men  using an anonymous data from the Baltimore longitudinal study of aging.
	In the study, 1590 healthy male volunteers were scheduled to visit the Gerontology
		Research Center bi-annually. Systolic blood pressure and current age were recorded during each visit. The design of the data is sparse and irregular, as many visits were missed by participants or not on the schedule; see \cite{Pearson1997} 
		for more details. Our study aims to predict the average systolic blood pressure in late middle age, between 55 and 60, based on the blood pressure trajectory between age 45 and 55. By excluding subjects with at most one visit between age 45 and 55 and no visit between 55 and 60, we obtain a subset of the data with $n=323$ subjects and on average 4.2 visits per subject, shown in the right panel of Figure
			\ref{fig:data}. The average of estimated intrinsic dimensions is 2.4 with a standard error 0.069, while the average number of principal components for functional linear regression is 4 with a 
			standard error 2.01. Based on Table \ref{tab:spectrometric}, our method outperforms others significantly. 
	
	\begin{table}
		\def~{\hphantom{0}}
		\tbl{Results for real data anslysis}{%
			\begin{tabular}{cccccccc}
				& FLR & FNW & FCE & FMO & FCM & MUL & {FREM}\tabularnewline
				MSP & $2.56\, {(0.43)}$ & $2.42\, {(0.33)}$ & $1.97\, {(0.35)}$ & $2.66\, {(0.46)}$ & $2.82\, {(0.45)}$ & $2.31\, {(0.35)}$ & $1.06\, {(0.34)}$\tabularnewline
				DTI & $1.14\, {(0.09)}$ & $1.28\, {(0.12)}$ & $1.36\, {(0.13)}$ & $1.78\, {(0.16)}$ & $1.25\, {(0.14)}$ & $1.33\, {(0.13)}$ & $0.96\, {(0.09)}$
				\tabularnewline
				SBP & $1.34\, {(0.18)}$ & $1.57\, {(0.17)}$ & $1.64\, {(0.16)}$ & $2.33\, {(0.26)}$ & $1.68\, {(0.19)}$ & $1.76\, {(0.17)}$ & $1.15\, {(0.11)}$
				\tabularnewline
		\end{tabular}}
		\label{tab:spectrometric}
		\begin{tabnote}
			FLR, functional linear regression; FNW, functional Nadaraya--Watson smoothing; FCE, functional conditional expectation; FMO, functional mode, FCM, functional conditional median; MUL,  multi-method; MSP, meat spectrometric data; FREM, the proposed functional regression on manifold; DTI, diffusion tensor imaging data; SBP, systolic blood pressure data. The numbers outside of parentheses are the Monte Carlo average of root mean square error based on 100 independent simulation replicates, and the numbers in parentheses are the corresponding standard error. The results for the  diffusion tensor imaging data and systolic blood pressure data are scaled by 0.1 for visualization.
		\end{tabnote}
	\end{table}

\section*{Acknowledgement}

Fang Yao's research is partially supported by National Natural Science Foundation of China (Key Grant 11931001 and General Grant 
11871080), and the Key Laboratory of Mathematical Economics and Quantitative Finance (Peking University), Ministry of Education.

\section*{Supplementary material}
Additional details and simulation studies for sparse functional data, the proofs of main theorems, auxiliary results, and technical lemmas with proofs are collected in an online Supplementary Material for space economy.

	\bibliographystyle{biometrika}
	\bibliography{fmanifold,fda,other}
	
	\includepdf[page=-]{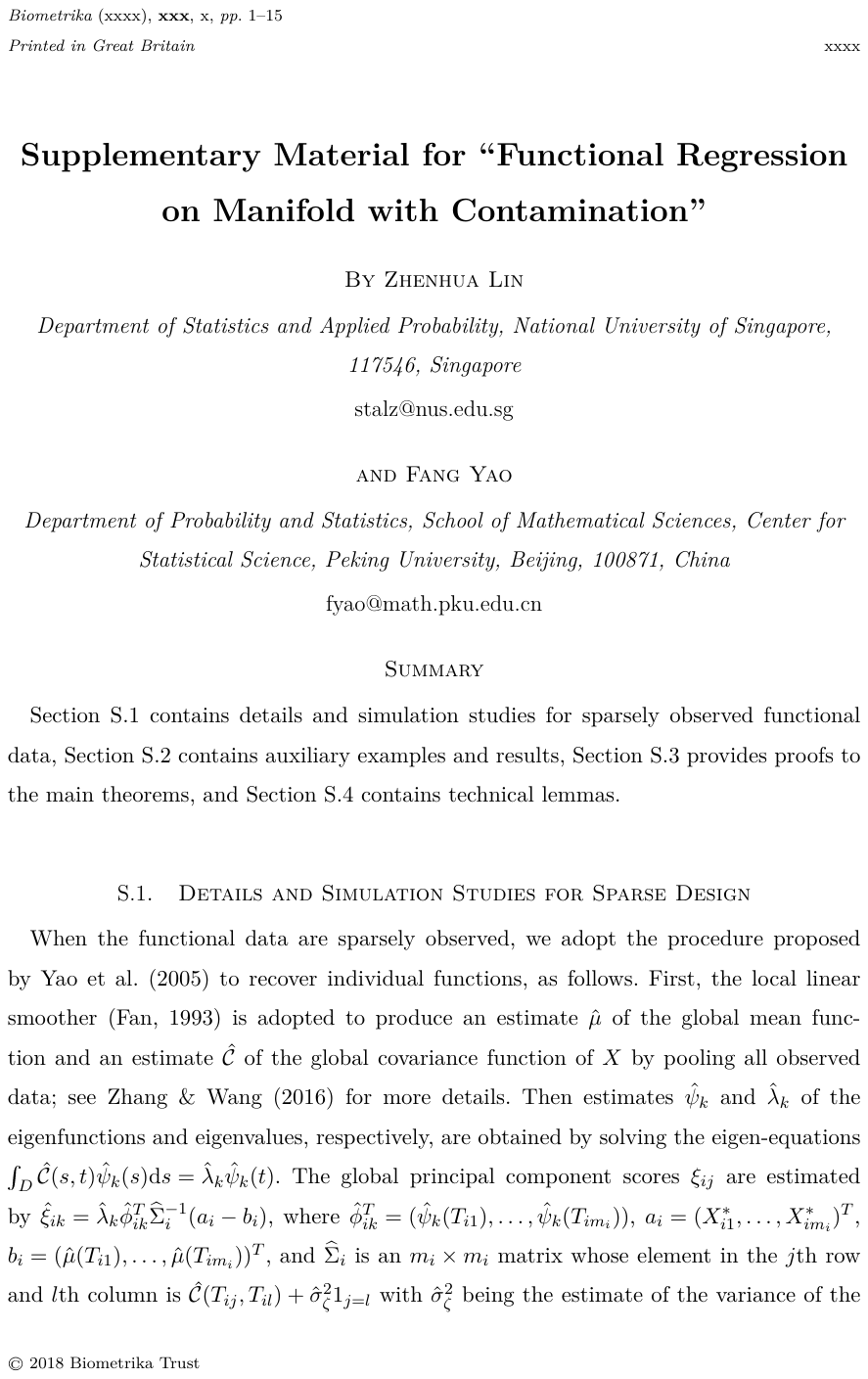}
	
\end{document}